\documentstyle[a4,12pt]{article}
\begin{document}
\begin{figure}[ht]
\vskip 4.0cm
\end{figure}
\begin{center}
\Large
Haldane Gap and Hidden Order in the $S=2$
Antiferromagnetic Quantum Spin Chain \normalsize
\vskip 2.0cm
\large
Ulrich Schollw\"{o}ck and Thierry Jolic\oe ur \\ \normalsize
\vskip 0.3cm
Service de Physique Th\'{e}orique \\
Commissariat \`{a} l'Energie Atomique, Saclay \\ 91191 Gif-sur-Yvette CEDEX
\\
France\\
\vskip 0.5cm
January 23, 1995
\end{center}
\vskip 1.0cm
We have investigated Haldane's conjecture for the S=2 isotropic
antiferromagnetic quantum spin chain with nearest-neighbor exchange J.
Using a
density matrix renormalization group algorithm for chains up  to L=350
spins, we
find in the thermodynamic limit a finite gap of $\Delta = 0.085(5)$J
and a finite spin-spin correlation length $\xi = 49(1)$ lattice spacings.
We
establish the ground state energy per bond to be E$_{0}=-4.761248(1)$J.
We show that the ground state has a hidden topological order that is
revealed in a
nonlocal string correlation function. This means that the physics of the
S=2
chain can be captured by a valence-bond solid description. We also observe
effective
free spin-1 states at the ends of an open S=2 chain.
SPhT/95-007 \\
cond-mat/9501115
\pagestyle{empty}
\newpage
\pagestyle{plain}
\pagenumbering{arabic}
Since Haldane's conjecture\cite{Haldane 83} that the physical properties of
(isotropic) antiferromagnetic quantum spin chains depend crucially on
whether the
spin is integer or half-integer, his prediction of a ground state with a
finite
spin-spin correlation length  and a finite gap to spin excitations
for integer spins has been studied in numerous papers. Numerical methods
have served to establish quantitative results where analytical work had to
rely on
often uncontrolled approximations.

Both the existence of a finite gap between a ground state singlet and
excited
states and of a finite correlation length is well established for the S=1
isotropic
antiferromagnetic quantum spin chain\cite{numerics}. Recent estimates of
the Haldane
gap are discussing the fifth digit: $\Delta = 0.41050(2)$\cite{White 93}
from the Density Matrix
Renormalization Group (DMRG) algorithm\cite{White 92}, and $\Delta =
0.41049(2)$ \cite{Golinelli 94} from extrapolation of
exact diagonalization results for chains up to length L=22. Similarily,
the
correlation length is $\xi = 6.03(1)$ \cite{White 93,corlens1}. The
spin
wave velocity is then found to be $c=\Delta\xi = 2.475(5)$, to be compared
to the
semiclassical value $c=2S=2$.

In this work we have established numerical values for the respective
quantities
for the S=2 isotropic antiferromagnetic spin chain with nearest-neighbor
exchange using the DMRG algorithm. We show that Haldane's conjecture is
obeyed in
this case: the Haldane gap also exists for
S=2, further supporting the case  that the gap exists for all integer
spins.
In addition our results show that a valence-bond-solid picture\cite{Affleck
87} is
valid for the S=2 spin chain.

The investigation of the $S=2$ case is hindered by the fact that one
expects $\Delta_{S=2}\ll \Delta_{S=1}$ and $\xi_{S=2} \gg \xi_{S=1}$. This
belief
is due to the asymptotic form $\Delta_S\approx S^2 \exp(-\pi
S)$\cite{Haldane 83}.
One may approximatively expect $\Delta_{S=2} \approx 0.07$ and
$\xi_{S=2} \approx 70$.
The much larger correlation length makes finite size extrapolations
feasible only
for much longer chains than in the S=1 case; furthermore, the inherent
statistical or systematical imprecisions of all numerical methods become
more
worrisome due to the small size of the gap. At the same time, the number of
states
per site rises from 3 to 5, greatly reducing the length of numerically
tractable
chains. At the present time, there are only a few preliminary
estimates\cite{Deisz
93,Japs 93} in the literature.

The DMRG algorithm allows to treat very long chains while retaining good
precision
for the energies and expectation values by an iterative truncation of
the Hilbert space down to M basis states. For a detailed discussion of the
method we refer to \cite{White 93, White 92}. In the following we have
investigated chains up to a length of L=350, which we will find to be about
$7\xi$.
This allows for sensible finite size extrapolations. At the same time, we
have
retained up to M=210 states, which allows for reasonable accuracy.
As the DMRG is much more precise for open than for closed chains,
we have considered only open chains.

To calculate the gap, we consider an open S=2 chain with a spin-1 at
each end
($S_1=S_L=1$; $S_i=2$ otherwise):
\begin{equation}
H = J_{end}\, \mbox{\bf S}_1\cdot\mbox{\bf S}_2 + J \sum_{i=2}^{L-2}
\mbox{\bf
S}_i\cdot\mbox{\bf S}_{i+1} + J_{end}\, \mbox{\bf S}_{L-1}\cdot\mbox{\bf
S}_L .
\end{equation}
We set $J_{end}=J$ and the results will be given in units of $J$ from now
on.

Throughout this study we are guided by the so-called valence-bond-solid
(VBS)
wavefunctions\cite{Affleck 87}. They are exact ground states of very
special spin
Hamiltonians (VBS hereafter). For S=1 it is well established that the
simplest
Heisenberg Hamiltonian\cite{Sorensen 93} shares a similar
physics with the VBS
Hamiltonian. These VBS ground states have a finite correlation length and
there is a gap to spin excitations.
 From these VBS Hamiltonians it is known that the ground state of the open
VBS-chain with integer spin $S$ is exactly
$(S+1)^2$-fold degenerate\cite{Affleck 87}, because the chain ends act like
free
$S/2$ spins. For a generic ``Haldane-like" Hamiltonian we expect this
manifold of
states to be split due to interactions between end spins.  We couple a
spin-$S/2$ to
each end with an antiferromagnetic coupling $J_{end} > 0$ (we chose
$J_{end}=1$)
to select a state in this manifold.

We calculated the gap for L=60,90,120,150,210,270, as the difference
between
the total energies of the lowest energy states in the $S^z_{total}=0$ and
$S^z_{total}=1$ subspaces. We have kept M=210 states in the DMRG.
White has given a procedure\cite{White 92}, which allows to ameliorate
iteratively
the results found for the total energies of a chain of fixed length L.
For short
chains, the DMRG algorithm truncates less well than for long chains; after
the chain
has grown to full length L, the results can be used to ameliorate the
truncations
at the start of the algorithm. We do not apply this iteration to the ground
state,
but iterate once for the excited state. From test calculations we found
that
iterating (further) does not substantially increase the precision obtained
or
influence our extrapolation. We extrapolate the results first for fixed L
to M=$\infty$, and then these extrapolated values
are used to investigate the
thermodynamic limit L=$\infty$.

We have extrapolated the energies to M=$\infty$ and deduced the gap values
by
subtraction. We find that both for the (uniterated) ground state energy and
(iterated) excitation energy the deviation from the asymptotic value is
proportional to $(1-\sum\lambda_{i})$, where $\lambda_i$ are the eigenvalues
of
the states of the  density matrix that are kept at the last iteration. The
proportionality coefficient is in both cases of  order 10 (much larger for
the
uniterated excitation energy), the truncation error
$1-\sum\lambda_{i}$ for M=210 between $3.7\times 10^{-8}$ and $9.4\times
10^{-8}$ for
the ground state energy and $6.7\times10^{-8}$ and $5.3\times 10^{-7}$ for
the
iterated excitation energy for
L between $60$ and $210$. The precision of the gap is in reality better
than
estimated, because both energies are overestimated due to the variational
character
of the DMRG, and there will be partial cancellation of the errors.

The ground state energy results allow us to calculate the ground state
energy
per site E$_0$. We find that for L$\approx$120, for each M$\leq 210$
considered
the ground state energy per site has converged to a precision of $10^{-6}$.
These
values we then extrapolate in M, again finding a linear dependence of the
error on
the truncation error. We conclude that E$_0 = -4.761248(1)$.

We can extrapolate our results for $90 \leq$ L $\leq 270$ to L=$\infty$ by
fitting
an L$^{-1}$-law (figure 1) and obtain a gap estimate of $0.081(1)$ for
L=$\infty$.
We note that deviations from perfect linear behaviour are extremely small
in the
range of chain length we have considered.
However it is important to note that the asymptotic behaviour of the gap in
an open
chain is expected to be $1/L^2$: the massive quasiparticles have dispersion

$E_k =\sqrt{\Delta^2_{\infty}+c^2 (k-\pi)^2}\approx
\Delta_{\infty}+c^2(k-\pi)^2/2\Delta_{\infty}$
near the
bottom of the band. On an open chain the quasiparticle cannot stay in a
stationary
state with $k=\pi$ but instead $k-\pi \approx \pi/L$ as a particle in a box
since
there is no translational invariance. This means that the lowest excited
state has
a gap $\Delta_{\infty} +c^2 \pi^2 /2\Delta_{\infty} L^2 +O(1/L^3)$.
This behaviour has not
yet been
reached for L=270. Thus there must be a crossover point to parabolic
behaviour of
$\Delta(L)$. As a consequence we can obtain an upper bound on
the gap
by assuming that the parabolic behaviour sets in immediately beyond L=270.
We
match a parabolic curve $\Delta (L) = \Delta_{\infty} + aL^{-2}$ to our
extrapolated
linear gap curve at L=270 such that up to the first derivative
the two
regimes meet continuously. This leads to $\Delta_{\infty} =0.090$ (and
$a\approx$620). We have also estimated the correlation length (see below)
to be
$\approx$50 and thus $c=\xi\Delta_{\infty}\approx 4.2$. This gives an estimate
of
the
coefficient of the $1/L^2$ parabolic term. Using this estimate
($a_0\approx$1024) we
now say that there is a crossover length $L_0$ at which asymptotic
behaviour
$\Delta(L) =\Delta_{\infty} +a_0/L^2$ sets in. Matching the parabolic curve to
the
straight line requires $\Delta_{\infty}=0.085$ and $L_0\approx 450$. This is a
perfectly
consistent set of results. We thus choose as a central value for the gap
$\Delta_{\infty}
=0.085$ the lower bound being fixed by the linear fit of our data and the
upper
bound being $0.09$. We quote our final result as $\Delta =0.085(5)$. A
gapless
state is excluded beyond any doubt.

To calculate the correlation length, we partially lift the
ground state degeneracy of the open chain by adding a spin-1 on {\em one}
end of the
S=2 chain ($S_L=1$; $S_i=2$ otherwise):
\begin{equation}
H = J
\sum_{i=1}^{L-2} \mbox{\bf S}_i\cdot\mbox{\bf S}_{i+1} + J_{end} \mbox{\bf
S}_{L-1}
\cdot\mbox{\bf S}_L .
\end{equation}
The ground state will then be a spin-1 triplet; at the spin-2 end of
the chain (position 1) one expects an effective free spin-1. If we consider
the
ground state with $S^z_{total}=1$, it is expected that $\langle S^{z}_{i}
\rangle$
decays purely exponentially from a value around $\pm 1$:
$\langle S^z_{i} \rangle \propto (-1)^{i-1} \exp (-(i-1)/\xi).$
{}From this equation the correlation length $\xi$ can be obtained much more
precisely than from $\langle \mbox{\bf S}_i \cdot\mbox{\bf S}_j\rangle$, as
it is
derived from a one-point correlation, where the systematic errors of the
DMRG tend
to cancel partially, whereas they build up for the two-point correlations,
as
correlated states in each half-chain are systematically neglected at the
same time,
underestimating $\xi$.

Because of the slow increase of precision with M for S=2, we have to
perform an
extrapolation of $\xi$ with respect to M. At the same time we consider
chains of
length L=270 and repeat some calculations for L=350, to study possible
finite
size effects. We find that those are minor, as are effects of the choice of
$J_{end}$, and concentrate on the extrapolation in M. In figure 2,
we give
$\ln |\langle S^z_i\rangle|$ vs.\ $i$. The generic behaviour is the same
for all M:
After the expected decay of $S^z$, there is a small increase towards the
other end
of the chain (greatly exaggerated by the logarithmic scale). With
increasing M,
the minimum shifts to the right, the spin expectation value at the right
end of the
chain is greatly reduced, the decay becomes more exponential and approaches
a
limiting curve which we take to be the M=$\infty$ result. To derive $\xi$,
we give
in figure 3 the local decay length $2/(\ln |S^z_i| - \ln |S^z_{i+2}|)$
(averaged over two
neighbouring sites to reduce odd-even site oscillations).
One sees that at the spin-2 end of
the chain,
the correlation length is first rather short, but saturates to its bulk
value (the
shoulder). For M=180, the
convergence of $\xi$ for M$\rightarrow\infty$ can be well established, and
we find a
correlation length $\xi = 49(1)$ in a system which has thus length $L
\approx
5.5\xi$, which is consistent with minor finite size effects.
At the same time, we find that $\langle S^z_1\rangle$ converges towards
$1.13(1)$,
supporting the
picture of an effective free spin-1 at the end of the chain.

Furthermore, we have measured the following nonlocal string correlation:
\begin{equation}
G(n, m) =\langle S_n^{\alpha} \exp (i{\pi\over 2} \sum_{i=n+1}^{m-1}
S_i^\alpha )
S_m^\alpha\rangle .
\end{equation}
For the S=2 VBS-chain $G(n,m) \rightarrow -1$ for $|n-m| \rightarrow
\infty$, revealing a hidden topological order. Investigating isotropic
Heisenberg S=2 chains up to
L=270 and keeping up to M=150 states, we find that $G(n,m)$
saturates to
$-0.726(2)$ for $|n-m| \rightarrow
\infty$ indicative of hidden topological long-range order in the
Heisenberg S=2 chain\cite{Scholl 95}. These results imply that the physics
of the Heisenberg S=2 chain can be captured by a VBS description.

For the isotropic AFM quantum spin-2 chain we find from $\Delta =
0.085(5)$ and
$\xi = 49(1)$ a spin wave velocity $c=4.2(3)$. This result indicates that,
as
expected, S=2 is more classical than S=1. We have also shown that the
S=2
spin chain has hidden long-range order as in the VBS model.
 There is excellent evidence that the S=1 Haldane gap is realized in the
compound NENP\cite{jpr}. It is an open and interesting question to find an
experimental system with S=2 spins to check our findings.

It is a pleasure to thank N. Elstner, O. Golinelli and R. Lacaze for
valuable
discussions. The calculations were carried out on the CRAY C94 of the CEA.

\section{Figure Captions}

{\bf Figure 1}: Gap vs.\ $L^{-1}$ for L=90, L=120, L=150, L=210 and
L=270
for different M, the number of kept states. Full squares:
extrapolated values for M$\rightarrow \infty$. Solid line: extrapolation
of the gap
values to L$\rightarrow \infty$, yielding a lower bound.
The dashed line is an upper bound on gap
behaviour
which is obtained by assuming that the asymptotic form
$\Delta(L)=\Delta_\infty
+a/L^2$ starts at L=270. Since we have {\it no} evidence for such a
behaviour, the
dashed curve should be a strict upper bound on the gap.

{\bf Figure 2}: $\ln |\langle S^z_i\rangle |$ vs.\ site $i$ for an open
S=2 chain
with a spin-1 attached at the right end ($J_{end}=1$). Note that for
increased
chain length the straight-line fit for the slope is possible for larger
$i$, while
the slope itself remains virtually unchanged. The expectation values for
the right
end of the chain have obviously not converged yet, whereas the asymptotic
M$\rightarrow \infty$ slope value can be determined.

{\bf Figure 3}: Two-site average of the local correlation
length $2/(\ln |\langle S^z_i \rangle | -
\ln
|\langle S^z_{i+2} \rangle |)$ vs.\ site $i$ for $i<125$ for the
same chain as in Figure 2. The deviations
from the
shoulder in the center of the diagram are finite size and finite precision
effects
(for the right end).

\end{document}